\documentclass{aastex63}

\received{June 1, 2019}
\revised{January 10, 2019}
\accepted{\today}
\submitjournal{APJ}

\graphicspath{{./}{figures/Results Plots/}}
\usepackage{url}
\usepackage{amssymb}
\usepackage{dsfont}
\usepackage{amsmath}
\usepackage{float}

\begin{document}

\title{Predicting the redshift of {\color{black}$\gamma$-ray loud} AGN using supervised machine learning}

\correspondingauthor{Maria Giovanna Dainotti}
\email{maria.dainotti@nao.ac.jp}

\author{Maria Giovanna Dainotti}
\affiliation{National Astronomical Observatory of Japan, Mitaka}
\affiliation{Space Science Institute, 4750 Walnut St, Suite 205, Boulder,CO,80301,USA}

\author{Malgorzata Bogdan}
\affiliation{Department of Mathematics, University of Wroclaw, Poland}
\affiliation{Department of Statistics, Lund University, Sweden}

\author{Aditya Narendra}
\affiliation{Jagiellonian University, Poland}

\author{Spencer James Gibson}
\affiliation{Carnegie Mellon University, USA}

\author{Blazej Miasojedow}
\affiliation{Faculty of Mathematics, Informatics and Mechanics, University of Warsaw, Poland}

\author{Ioannis Liodakis}
\affiliation{Finnish Center for Astronomy with ESO (FINCA), University of Turku, Finland}

\author{Agnieszka Pollo}
\affiliation{Astronomical Observatory of Jagiellonian University, Krakow}
\affiliation{National Centre for Nuclear Research, Warsaw}

\author{Trevor Nelson}
\affiliation{University of Massachusetts at Amherst, Massachusetts, USA}

\author{Kamil Wozniak}
\affiliation{AGH University of Science and Technology, Krakow}
\author{Zooey Nguyen}

\affiliation{Faculty of Astronomy, University of California, Los Angeles, California, USA}

\author{Johan Larrson}
\affiliation{Department of Statistics, Lund University, Sweden}





\begin{abstract}
AGNs are very powerful galaxies characterized by extremely bright emissions coming out from their central massive black holes. Knowing the redshifts of AGNs provides us with an opportunity to determine their distance to investigate important astrophysical problems such as the evolution of the early stars, their formation along with the structure of early galaxies. The redshift determination is challenging, because it requires detailed follow-up of multi-wavelength observations, often involving various astronomical facilities. Here, we employ machine learning algorithms to estimate redshifts from the observed $\gamma$-ray properties and photometric data of 
{$\gamma$-ray loud} AGN from the Fourth Fermi-LAT Catalog. The prediction is obtained with the Superlearner algorithm, using LASSO selected set of predictors.   
We obtain a tight correlation, with a Pearson Correlation Coefficient of 
{71.3\% between the inferred and the observed redshifts, an average $\Delta$z$_{norm}$ = 11.6$\times 10^{-4}$. We stress that notwithstanding the small sample of $\gamma$-ray loud AGNs, we obtain a reliable predictive model using Superlearner, which is an ensemble of several machine learning models.}
\end{abstract}



\section{\textbf{Introduction}} \label{sec:intro}


Active Galactic Nuclei (AGN) with jets are the dominant class of objects when it comes to high-latitude ($|b|>10$) extragalactic $\gamma$-ray sources \citep{2020ApJS..247...33A}. 
The {\it Fermi} $\gamma$-ray space telescope has detected more than 2863 such $\gamma$-ray AGNs, the majority of which ($>98\%$) are blazars: AGN with their jets pointed towards our line of sight. 
Blazars are denoted by the equivalent width of resonant emission lines in their optical spectra. Sources with broad emission lines are classified as Flat Spectrum Radio Quasars (FSRQs), whereas sources with weak or no emission lines are classified as BL Lacertae objects (BLLs). 
Measuring the redshift (z) of blazars has been a cumbersome and observationally expensive endeavor. The situation is further complicated by the absence of emission lines in the most numerous class of $\gamma$-ray loud blazars, i.e., BL Lacs. 
As a result, out of the 2863 sources of the Fourth AGN {\it Fermi}-LAT catalog (4LAC, \cite{2020ApJ...892..105A}), only 1591 have redshift estimates, ranging from $z=[0,3]$, but mostly concentrate below $z=2$. 
$\gamma$-Ray loud blazars with redshift estimates are relevant for our comprehension of the origin of the Extragalactic Background Light (EBL), which in turn let us probe the cosmic evolution of blazars 
(e.g., \cite{singal2012flux}, \cite{singal2014gamma}, \cite{singal2015determination}, \cite{singal2013flat}, \cite{chiang1995evolution}, \cite{ackermann2015multiwavelength}, \cite{singal2013cosmological}
\citealp{2020AAS...23540506M}), the intergalactic magnetic field (e.g., \citealp{2013MNRAS.432.3485V}), star formation rate history of our universe (e.g., \citealp{2018Sci...362.1031F}), as well as constrain cosmological parameters (e.g., \citealp{2019ApJ...885..137D}). The difficulty in spectroscopically measuring redshift in a significant fraction of BL Lacs and the importance of identifying high-$z$ blazars has led to the development of photometric estimation techniques ({\bf photo-z}, e.g., 
\citealp{2017ApJ...834...41K,2018ApJ...859...80K,2020ApJ...898...18R,carrasco2015photometric,krakowski2016machine,nakoneczny2019catalog} 
). However, works using such methods typically produce redshift estimates for only $\sim6-13\%$ of their sample, making alternative methods necessary.
{ Machine learning (ML) methods for obtaining photo-z estimates for AGN are becoming increasingly important in the era of big data Astronomy (e.g., \citealp{DIsanto2018,brescia2013photometric,Brescia2019,ilbert2008cosmos,hildebrandt2010phat}). Here we focus on the $\gamma$-ray emitting AGN population in the 4LAC.}

In the current literature, multiple works exist which focus on extracting reliable photometric redshift of AGNs
\citep{cavuoti2014photometric,fotopoulou2018cpz,logan2020unsupervised,yang2017quasar,zhang2019machine,curran2020qso,  nakoneczny2020photometric,pasquet2018deep,jones2017analysis}.
{In the current blazar literature, a lot of effort has also been placed in classifying blazars of uncertain type (e.g., \citealp{Chiaro2016,kang2019evaluating}) and unidentified {\it Fermi} objects (e.g., \citealp{Liodakis2019}). Although these papers convey useful information about the algorithms that work well for classifying blazars, so far no analysis has been performed regarding the prediction of the redshifts of $\gamma$-ray 
loud blazars. }Thus, we will tackle this problem by using machine and statistical learning algorithms.
We apply multiple ML algorithms, such as LASSO (Least Absolute Shrinkage and Selection Operator), XGBoost (Extreme Gradient boosting), RandomForest, and BayesGLM (Bayesian generalized linear model). 
We follow the approach used in \cite{dainotti2019gamma}, where some of us used the SuperLearner package to aggregate the results from multiple algorithms and predict the redshifts of $\gamma$-ray bursts.

The results of this study increases the number of blazars with inferred redshifts considerably so that we can finally obtain a more complete sample of $\gamma$-ray loud AGNs. As a result, this work will enable the solving of some crucial questions on the luminosity function and density evolution of $\gamma$-ray loud AGNs.

In Section \ref{sec:sample}, we discuss the data and predictors used. In Section \ref{sec:methods},
we outline the ML methods used,
the selection of the best predictors and algorithms, and the validation of our results. 
In Section \ref{sec:results}, we present the results obtained in this analysis.
In Section \ref{sec:conclusion}, we present our results and discuss future perspectives.

\section{\textbf{The sample}} \label{sec:sample}
\begin{figure}[H]
\centering
\includegraphics[width=\textwidth]{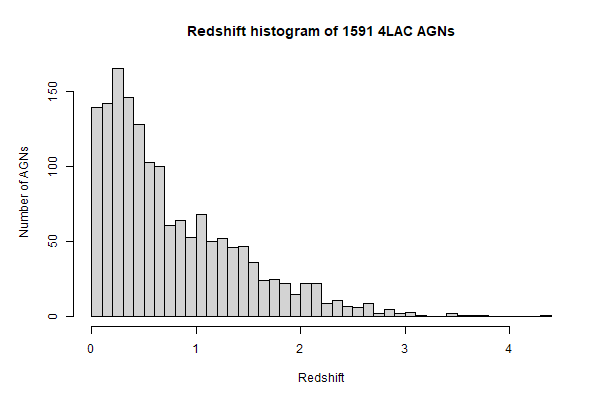}
\caption{The redshift distribution of the entire 4LAC catalog before selection cuts and outliers removal. }
\label{fig:red1}
\end{figure}

\begin{figure}[H]
    \centering
    \includegraphics[width=0.99\textwidth]{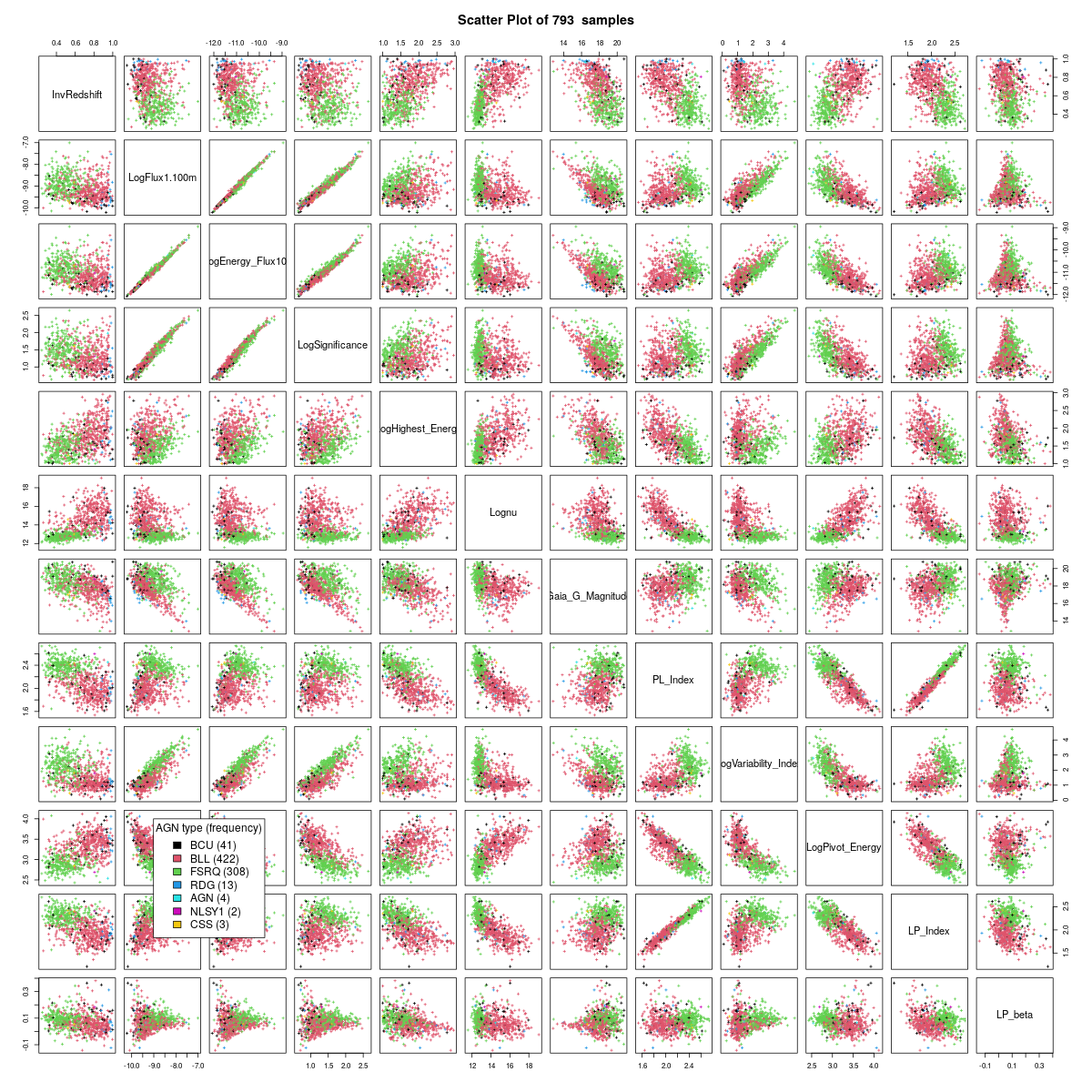}

    \caption{The full scatter matrix plot of all the variables defined above, before feature selection. Here the \textit{InvRedshift} denotes $\frac{1}{z+1}$ scaled data.}
    \label{fig:fullscatter}
\end{figure}

{\it Fermi}-LAT has been continuously monitoring the sky in the 50 MeV to 1~TeV range since 2008. The $\gamma$-ray properties used in this work are obtained from the 4LAC catalog \citep{2020ApJ...892..105A}.
{\bf It contains 2863 sources, 658 of which are FSRQs, 1067 are BL Lacs, 1074 are blazars of uncertain type, and the remaining 64 sources are classified as radio galaxies, Narrow line Seyferts (NLSY1), and other non-blazar AGNs. 
Out of the 2863 sources, 1591 have a measured redshift, whose distribution is shown in Fig. \ref{fig:red1}.}
{For completeness of the treatment we have included also non BL LAC and non FSRQs sources in the initial scatter matrix plot in Fig. \ref{fig:fullscatter} to show how the  variables in the sample is distributed. But, in the generalization set, we are predicting the redshift for only the BLLs. 
}

{Unfortunately, all of the 1591 $\gamma$-ray AGNs cannot be used for our model's training. A significant number of these $\gamma$-ray AGNs have incomplete observational data, meaning we face the problem of missing values in several parameters.}

Thus, we perform cuts in the data set to remove incomplete data points leaving us with 1169 $\gamma$-ray AGNs out of 2863. These consist of 661 BLLs, 309 FSRQs, 177 unclassified AGNs, and 22 AGNs belonging to other categories.
This set is split into training and generalization sets, the former consisting of the $\gamma$-ray AGNs that have observed spectroscopic redshift, while the latter consists of the $\gamma$-ray AGNs for which the redshift is not measured.
Our training set consists of 793 $\gamma$-ray AGNs, made up of 422 BLLs, 308 FSRQs, 41 unclassified, and 22 other category AGNs.
\textcolor{black}{The 22 other category $\gamma$-ray AGNs in our training set consisted of 2 NLSY1 sources, 3 Compact-Steep spectrum Radio Source (CSS) sources, 13 Radio Galaxies (RDG) sources, and 2 sources classified as non-blazar AGNs. They are shown in Fig. \ref{fig:fullscatter}. After we perform the cuts related to the missing data we are left with 730 $\gamma$-ray AGNs.}
Similarly, our generalization set consists of 376 $\gamma$-ray AGNs, of which are 239 BLLs, 1 FSRQ, and are 136 unclassified AGNs.
After we perform the cuts in the generalization set we are left with 239 BLLs.
\textcolor{black}{Due to their dominating presence, we perform our predictions only for BLLs, and remove the 136 uncategorized AGNs. But, in the scatter matrix plot of Fig. \ref{fig:Scatter}, we show in black the only  FSRQ from the generalization set. 
}

{BL Lacs and FSRQs can be very easily separated as we did when we have introduced in the Superlearner categorical variables. We here stress that this is an important point, because it means that the quality of the predictions will most probably differ, especially if the fractions of BL Lacs in the training sample and in the full population are very different. This is expected as we have already mentioned in the introduction that this could be the case because of the difficulty of obtaining their spectroscopic redshift. We also would like to stress that due to the paucity of the other classes the categorical variables have been limited to BLL and FSRQs.}





\begin{figure}[H]
    \centering
    \includegraphics[width=\textwidth]{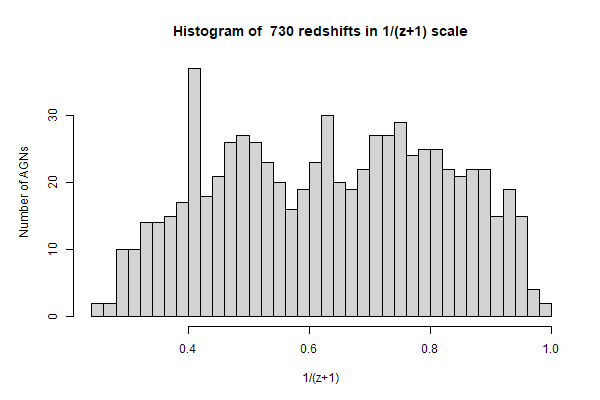}
    \caption{The histogram distribution of the redshift of our training set in 1/(z+1) scale}
    \label{fig:red2}
\end{figure}

Regarding the predictors, 4LAC contains 13 photometric variables along with the spectroscopic redshift and names of the AGNs.
It also includes the g-band magnitudes for individual sources from Gaia \citep{2010A&A...523A..48J}.
Some of the variables are used in their logarithmic form since they span over several orders of magnitude and we predict the redshift in the 
{scale of $\frac{1}{z+1}$ (see Fig. \ref{fig:red2}). } 
{Out of these 13 variables, we take into consideration 11. We exclude fractional variability due to the incompleteness of the AGN sample and Log$\nu$f$\nu$ as it is a second-order variable depending on Log$\nu$.
The definition and explanation for the 11 variables are given below.}

\begin{itemize}\label{best}
\item \textit{LogFlux} - Logarithm in the base of 10 of the integral photon flux, in photons/cm2/s, from 1 to 100 GeV.
\item \textit{LogEnergy\_Flux} - Logarithm in base of 10 of the energy flux, the units are in erg cm$^{-2}$ s$^{-1}$, in the 100 MeV - 100 GeV range obtained by the spectral fitting in this range. 
\item \textit{LogSignificance} - The source detection significance in Gaussian sigma units, on the range from 50 MeV to 1 TeV. 
\item \textit{LogVariability\_Index} - The sum of the log(likelihood) difference between the flux fitted in each time interval and the average flux over the 50 MeV to 1 TeV range. 

\item \textit{Log Highest\_Energy} - Measured in GeV, it is the energy of the highest energy photon detected for each source, selected from the lowest instrumental background noise data, with an associated probability of more than 95\%.
\item \textit{Log$\nu$} - Logarithm in base of 10 of the synchrotron peak frequency in the observer frame, measured in Hz.
\item \textit{PL\_Index} - It is the photon index when fitting the spectrum with a power law, in the energy range from 50 MeV to 1 TeV.
\item \textit{LogPivot\_Energy} - The energy, in MeV, at which the error in the differential photon flux is minimal, derived from the likelihood analysis in the range from 100 MeV - 1 TeV.
\item \textit{LP\_Index} - Photon index at pivot energy ($\alpha$) when fitting the spectrum (100 MeV to 1 TeV) with Log Parabola. 
\item \textit{LP\_$\beta$} - the spectral parameter ($\beta$) when fitting with Log Parabola spectrum from 50 MeV to 1 TeV.
\item \textit{Gaia\_G\_Magnitude} - Gaia Magnitude at the g-band provided by the 4LAC, taken from the Gaia Survey.


\end{itemize}

 

\section{\textbf{Methodology}} \label{sec:methods}

In this section, we describe, in detail, the methodology adopted for this study, from the description of the choice of the transformations adopted, the variable selection, the methods considered singularly, such as {Big LASSO} (a more reliable version of LASSO), {XGBoost}, {Random Forest} and {Bayes GLM}, to the Superlearner algorithm used to create the ensemble leading to the final prediction (see Sec. \ref{sec:SL}).

{The statistical parameters used in order to compare our results with those of others in the field are: Bias, $\sigma_{NMAD}$ (normalized median absolute deviation), Pearson correlation $r$, RMSE (root mean square error), and standard deviation ($\sigma$). We quote the measured values of these parameters for $\Delta z_{norm}$ and $\Delta z$
, where 
$\Delta z_{norm}$ = $\frac{z_{spec} - z_{pred}}{(1+z_{spec})}$ and  
$\Delta z$ = $z_{spec} - z_{pred}$
}.
{As shown in the scatter matrix of Fig. \ref{fig:fullscatter}, we can see the presence of multiple correlated variables such as PL\_Index and LP\_Index, LogEnergyFlux and LogFlux, and LogFlux and LogSignificance. Hence, we deploy a feature selection method such as LASSO which as a result naturally reduces the number of correlated variables, although it does not completely eliminate all of them.}

\begin{figure}[H]
    \centering

    \includegraphics[width=0.65\textwidth]{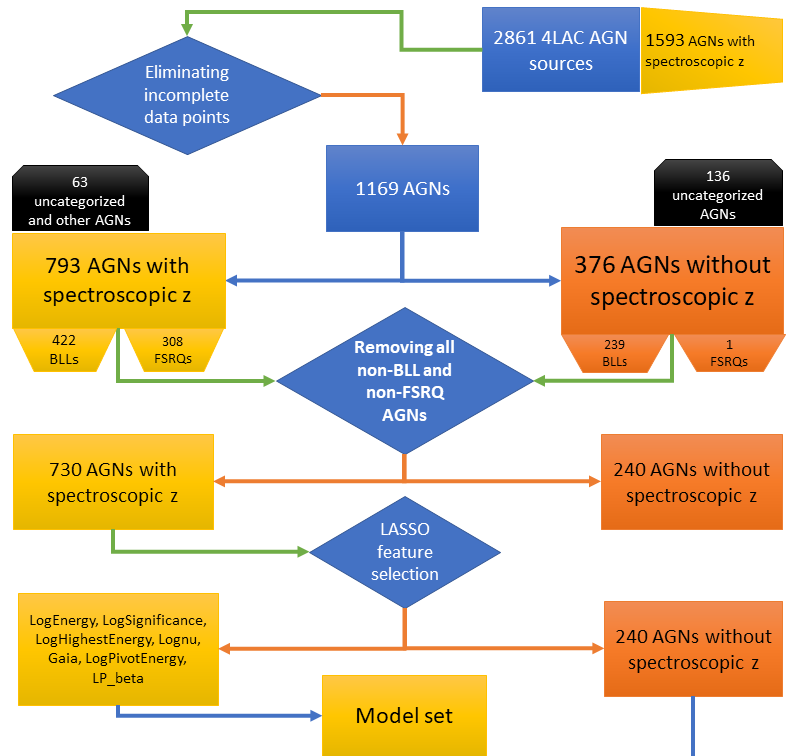}
    \includegraphics[width=0.65\textwidth]{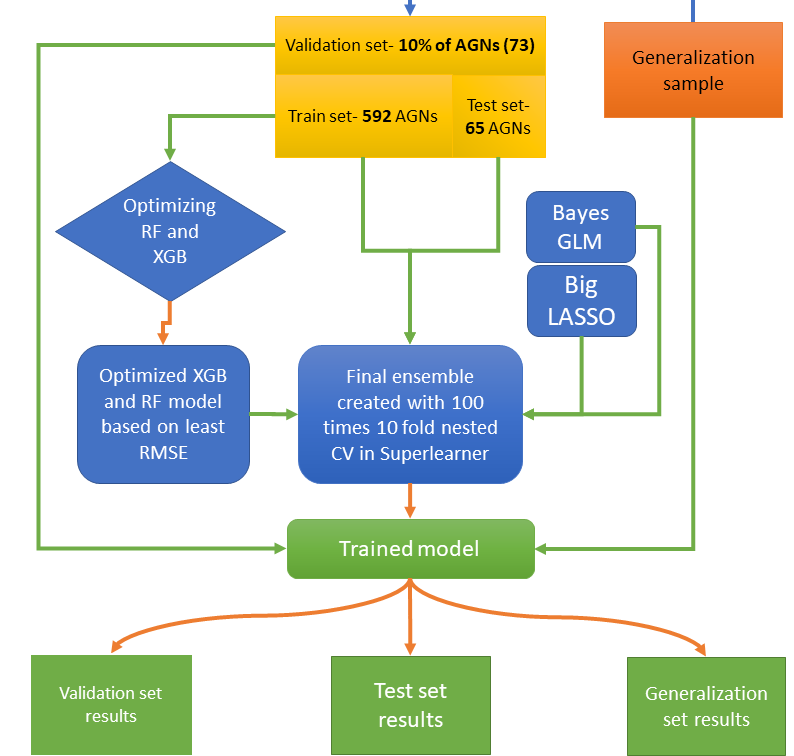}
   \caption{{ Methodology flowchart: the rectangular boxes represent data sets, the parallelograms the $\gamma$-ray AGN categories, the rhombus indicates functions performed, rounded rectangles indicate the ML algorithms used, the green lines show the direction of the input, orange lines the output, and blue lines indicate the splits and changes in the data set.
The color-coding indicates the following: yellow indicates the data with spectroscopic z, orange the ones without spectroscopic z, green the results and, blue indicates the intermediate steps or datasets.
}}
\label{fig:flowchart}
\end{figure}

{The procedure consists of mainly two parts, as presented in the flowchart in Fig. \ref{fig:flowchart}. The first steps are to clean our data source by eliminating data points with missing variables and then pruning our feature set with the use of the LASSO algorithm. 
After this, the variables obtained as the selected ones will be used to train our model.
{We split our data into train and test sets composed of 657 $\gamma$-ray AGNs, and the validation set composed of 73 $\gamma$-ray AGNs.  We divide the sample taking as the validation set the latest 10\% of the $\gamma$-ray AGN observed. This choice is the same as taking the validation set randomly since there is no preferential order in redshift when we choose the validation set. This is just for one test, but as we show in the Sec. \ref{sec:SL} we also apply the 10-fold cross-validation (hereafter called 10fCV) 100 times to avoid choosing a validation sample that may not be representative of the whole sample.}
{We will use Superlearner which includes the optimized XGBoost, Random forest, Bayes GLM, and Big LASSO. Details of such an optimization are mentioned in  Sec. \ref{sec:optimizing}. After training this ensemble on our data, we obtain our trained model, which leads us to the prediction on the redshifts.
}}

\subsection{\textbf{Feature selection}}\label{sec:FeatureSelection}
\begin{figure}[t]
    \centering
    \includegraphics[width=0.7\textwidth]{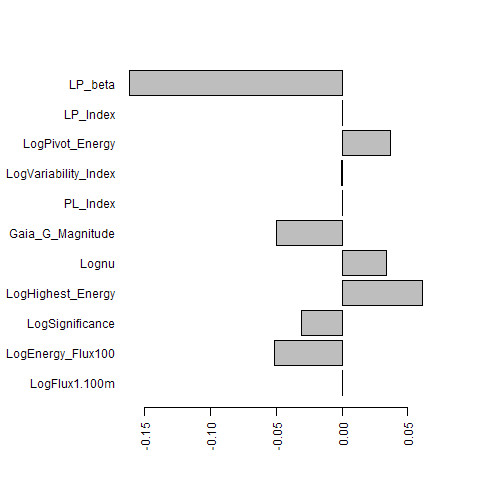}
    \caption{The coefficients assigned to the features by LASSO at the $\lambda$.1se value. We only keep the coefficient features $> 0$.}
    \label{fig:lassoF}
\end{figure}

\begin{figure}[H]
    \centering
    \includegraphics[width=1.0\textwidth]{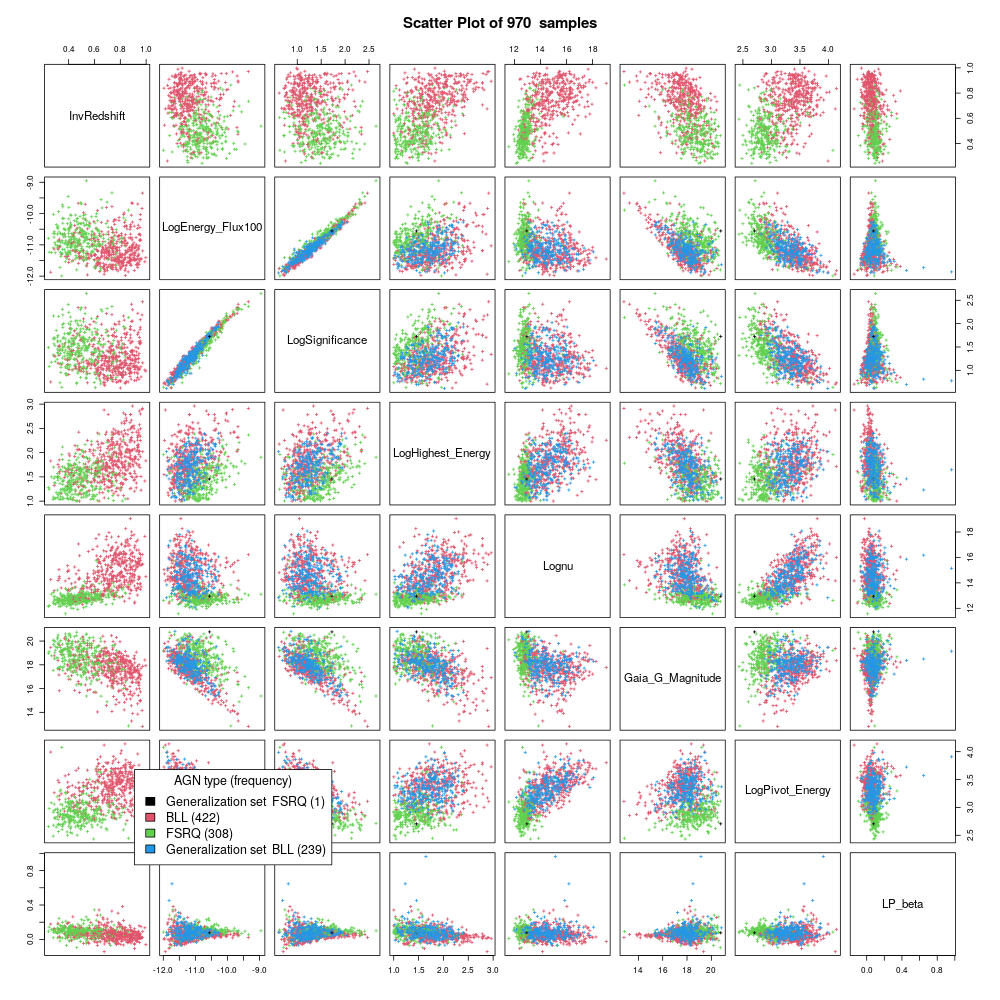}
    \caption{The full symmetric scatter matrix plot shows the response (in our case the \textit{InvRedshift}) and predictor variables. The different $\gamma$-ray AGN categories are color-coded according to the legend displayed on the plot. The values in the parenthesis indicate the number of $\gamma$-ray AGNs present in the data set. }
    \label{fig:Scatter}
\end{figure}

We apply the LASSO method to prune our features and obtain a more effective subset for redshift prediction.
The LASSO algorithm uses a shrinkage method for linear regression by requiring the $\ell^1$ norm (sum of the magnitude of all vectors in the given space)
of the solution vector to be less than or equal to a positive number known as the tuning parameter ($\lambda$). This penalization allows the model to select a subset of features and discards the rest by setting their coefficients to 0 \citep{TibshiraniLasso}.
The tuning parameter is responsible for deciding the shrinkage coefficient applied to the estimated vector. As a consequence, the model is easier to interpret with a smaller number of features and usually has a smaller prediction error than the full model. 
{The prediction error is the RMSE between the predicted and the observed redshifts, which is minimized during the one hundred times 10fCV training. As a measure of the prediction errors we quote the RMSE value, as well as the $\sigma_{NMAD}$.}
{For our analysis, we use the GLMNET function with the LASSO selection feature \citep{hastie2017extended,tibshirani2012strong}.
We pick the $\lambda.1se$ value, which is the maximum $\lambda$ value for which the error is within 1 standard deviation \citep{hastieTibs}} and its corresponding coefficients for the features.
{The coefficients assigned by LASSO to each of them are displayed in Fig. \ref{fig:lassoF} and we choose only the non-zero coefficient features.
To better visualize the parameter space of these features 
we plot them in the scatter matrix plot shown in Fig. \ref{fig:Scatter}
\textcolor{black}{, along with the generalization set}. 
}
LASSO feature selection shows that some of the variables that were strongly correlated are naturally eliminated, but we are still left with two correlated variables: LogEnergyFlux and LogSignificance.
{
This means that for LASSO both features are relevant. 
Since LogSignificance is providing the information on the detectability of the $\gamma$-ray AGN and this is relevant to the final prediction of the redshift; thus, we decided to retain it. 
On the other hand, from a statistical point of view, it is not necessary to remove correlated variables, since the aim here is to reach a greater accuracy on the prediction of the redshift. Nevertheless, we have shown in the Appendix (See Fig. \ref{fig:woLS}) that the results do not change at the level of $1\%$ for $\sigma_{NMAD}$ (Normalized Median Absolute Deviation), RMSE and Correlation when we consider to manually discard this variable. 
}

\textbf{ }



\begin{figure}
    \centering
    \includegraphics[width=\textwidth]{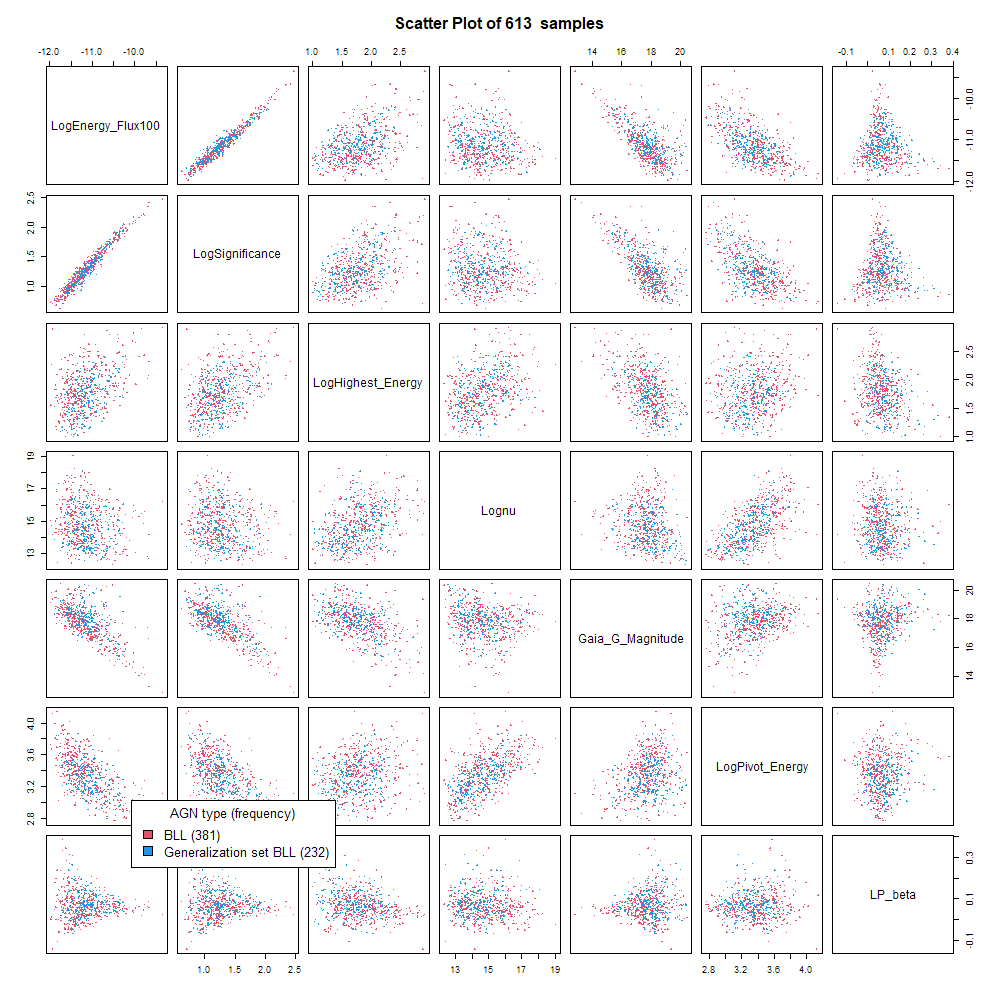}
    \caption{The scatter matrix plot for BLLs in the generalization and training set. The generalization set BLLs are shown in blue, while the training set BLLs are shown in red.}   
    \label{fig:overlap}
\end{figure}

{In addition, we clarify that we performed the analysis with both $\log_{10}(1+z)$ and $\frac{1}{z+1}$, the distribution of the latter shown in Fig. \ref{fig:red2}. The choice of transformation arises from the fact that 
the results related to the choice of $\frac{1}{z+1}$ present the smallest $\sigma_{NMAD}$  and smaller $\Delta z_{norm}$ (normalized variation in redshift), thus leading us to use this transformation. }

\subsection{The ML algorithms used in our analysis}\label{sec:ML}


By adopting an ML approach, we leverage the built-in algorithms that learn from the training set and we test out predictions on the test set.
We employ the trained models to predict the redshift of sources for which the redshift has not been measured.
These optimized methods are combined into an ensemble using the Superlearner package, providing us with a better prediction than any single algorithm.
The ML algorithms used here are summarized in the following itemized points:

\begin{itemize}
 
\item Regression trees build the predictor by partitioning the data based on the values of the independent variables and averaging the value of the dependent variables. Examples of regression trees are XGBoost and Random Forest. Indeed, both the XGBoost and Random Forest algorithms utilize multiple regression trees to increase their predictive power.
\item The Random Forest algorithm generates multiple independent regression trees and averages them to obtain a more accurate prediction \citep{breiman2001random,valencia2019can,green2019using,miller2015machine}.
An extremely difficult task is how to choose the optimal depth of such a tree, namely to decide which is the number of partition levels. In gradient boosting, the final predictor is built as a weighted sum of simple tree predictors. 
Compared to the Random Forest method, regression trees are not generated independently but built on each other using residuals from the previous step, until the culmination of trees forms a stronger regression model.
\item The XGBoost algorithm is an amelioration of the gradient boosting method \citep{chen2016xgboost,friedman2000additive,friedman2001greedy,friedman2002stochastic} and it also leverages poor predictors. It uses a more regularized model formalization to control overfitting, and thus give better performance.

\item Big LASSO is a computationally efficient implementation of the LASSO algorithm in R \citep{zeng2017biglasso}. 
The Big LASSO is an implementation that allows us to compute and 
analyze big multidimensional data sets quickly and efficiently.
\item 
{Bayes GLM is a bayesian inference of the generalized linear model. 
It determines the most likely estimate of the response variable (in our case the redshift) given the particular set of predictors and the prior distribution on the set of regression parameters (Maximum A Posteriori estimator, MAP).
It works on the Fisher principle: ``what value of the unknown parameter is \textit{most likely} to generate the observed data". 
BayesGLM method is more numerically and computationally stable as compared to normal GLM models. It employs a student-t prior distribution for the regression coefficients.}
{Then, given the observed data, the likelihood function for these parameters is calculated. The likelihood function and priors are combined to produce the posterior distributions from which we obtain the MAP estimators of the desired parameters \citep{birnbaum1962foundations,hastie1987generalized,hastie1990generalized,friedman2010regularization}}.

\end{itemize}

\subsection{Optimizing Algorithms}\label{sec:optimizing}

\begin{figure}[b]
    \centering

    \includegraphics[width=0.49\textwidth]{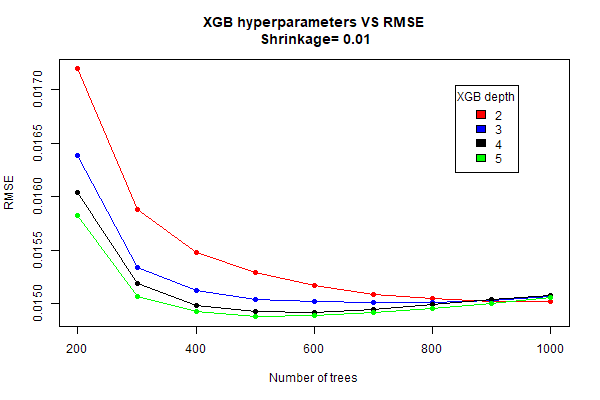}
    \includegraphics[width=0.49\textwidth]{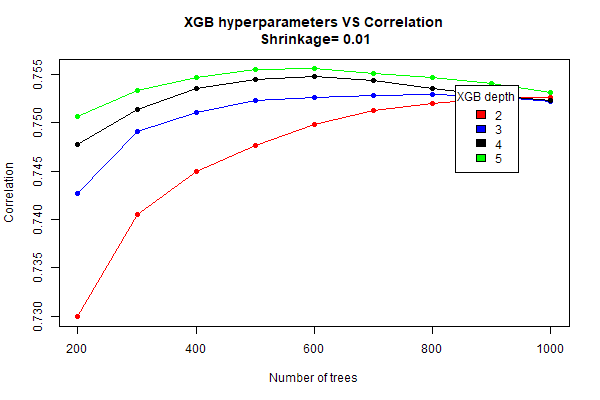}
    \includegraphics[width=0.49\textwidth]{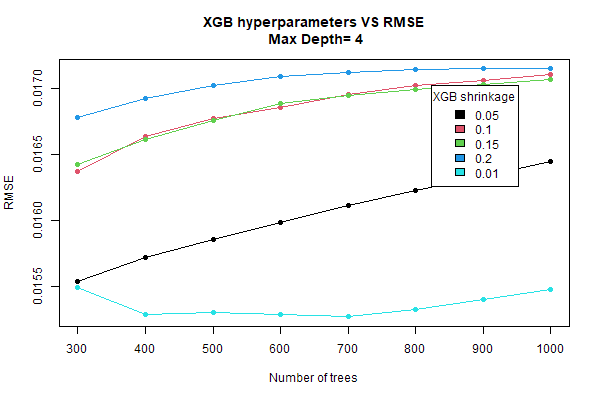}
    \includegraphics[width=0.49\textwidth]{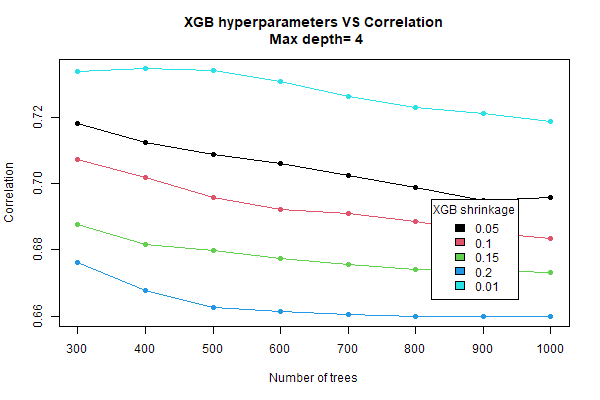}
    
    \caption{Variation of the RMSE and Correlation coefficient versus the number of trees, for different depths (upper panels) and shrinkage coefficients (lower panels). }
    \label{fig:xgbOpti}
\end{figure}

{It should be noted that these results are obtained after performing 10fCV on our data set. For the XGBoost algorithm, we have the option to vary the number of regression trees, the depth, and the learning rate (the so-called shrinkage coefficient, which shrinks the predictions of a tree to prevent over-fitting). We tune these to best fit our data without over or under-fitting. 
In Fig. \ref{fig:xgbOpti} top left and right panels show the variation of the root mean square error (RMSE) and correlation, respectively, related to the number of trees in the model. The RMSE and correlation minimize and maximize, respectively, at a depths of 5 and at a number of 500 trees. However, since depth 4 and 5 gives very similar results, to avoid the risk of over-fitting usually associated with a higher depth we choose a max depth of 4 and proceeded to test the model performance while varying the learning rate, see bottom panels of Fig. \ref{fig:xgbOpti}. The optimal learning rate in our case is 0.01.
In the left bottom panel of Fig. \ref{fig:xgbOpti}, we plot the RMSE variation, and on the right the Pearson Correlation coefficient (r). 
In summary, our final XGB optimized model consists of 500 trees, with a depth of 4 and a shrinkage coefficient of 0.01.}

\begin{figure}[H]
    \centering

    \includegraphics[width=0.48\textwidth]{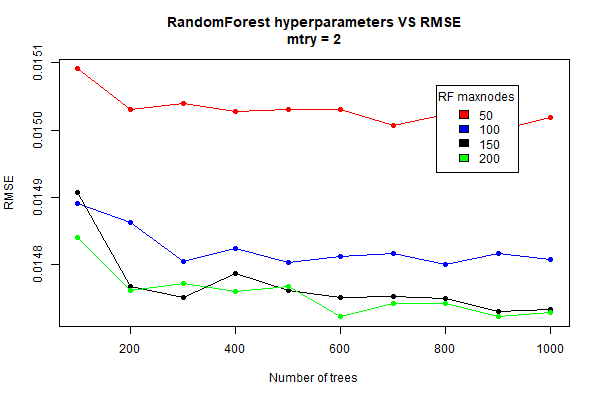}
    \includegraphics[width=0.48\textwidth]{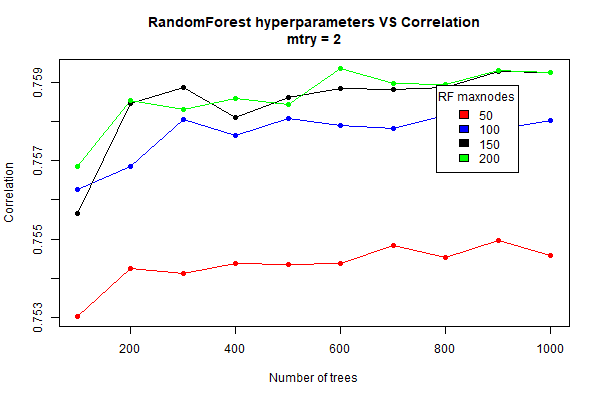}
    \includegraphics[width=0.48\textwidth]{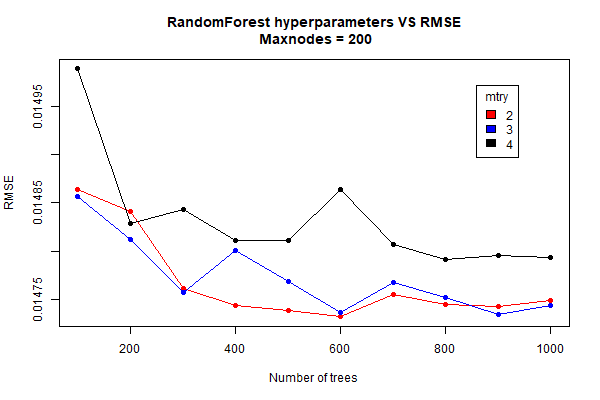}
    \includegraphics[width=0.48\textwidth]{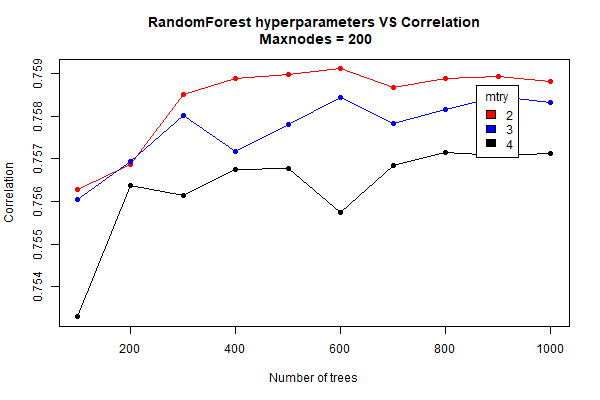}

    \caption{The panels show random forest optimization plots. The upper left and right panels present the RMSE and Correlation vs. the number of trees, respectively. This is performed with a fixed value of mtry=2 and different values of RF maxnodes=(50, 100, 150, 200) color-coded with red, blue, black, and green, respectively. The bottom left and right panels present the same plots as the upper panel, but with the fixed value of Maxnodes=200 and with mtry=2,3,4 indicated with red, blue and black, respectively.}
    \label{fig:rfOpti}
\end{figure}

{A similar analysis is performed for Random Forest as well. We tune the number of trees, depth, and the maximum number of nodes based on which model has the lowest RMSE and maximum correlation value.
We started with a default value for the number of variables that will be randomly sampled (from here on denoted as mtry), which is 2. We vary the number of trees and the maximum number of nodes. The RMSE and Correlation variation are shown in the top left and right plots of Fig. \ref{fig:rfOpti}, respectively. We observe that a value of 200 for maximum nodes gives the least RMSE and maximum correlation at 400 trees. 
Next, we keep the maxnode parameter constant and vary the mtry value from 2 to 4. The RMSE and Correlation plots are shown in the bottom panel of Fig. \ref{fig:rfOpti}. Among the different values of mtry tested, we see that mtry=2 gives us the best results in terms of the highest correlation coefficient and the smallest RMSE.} 
{Furthermore, the number of trees is selected to be 600, as this gives the second smallest RMSE, but since in this region we have contemporaneously also the plateau of the Correlation coefficient (see left bottom panel of Fig. \ref{fig:rfOpti}) 600 is the most favored value. In addition, when the RMSE is similar as in the 600 and 900 trees we prefer the smaller number of trees to prevent overfitting.}
{In the case of BayesGLM, there are no tuneable hyperparameters, as instead it is for XGBoost and RF. Instead, we specify a formula based on which the redshift is predicted. The formula used is a linear combination of all the features we consider:
}

\begin{equation}
    \frac{1}{z_{i}+1} = f (\sum K_{i})
\end{equation}
\\
{Here $K$ belongs to a set of features described in Sec. \ref{sec:FeatureSelection} and presented in Fig. \ref{fig:lassoF}, and $i$ denotes each $\gamma$-ray AGN in the training set which is used in the model fitting. }

{The Big LASSO algorithm is an extension of LASSO. Hence its optimization is done identically, i.e its $\lambda$ hyperparameter is tuned based on its internal CV such as to obtain the model with the least RMSE. As a result, there is no need for us to explicitly handle its optimization. }

{Since every ML method has its advantages in a given parameter space and in our case in different redshift ranges, we leverage each of the methods by using Superlearner, described in the next subsection.
}

\subsection{SuperLearner}\label{sec:SL}
{In our approach, we have three different types of sets: the training, the test, and the generalization sets.
The training set is used to train the model based on the observed variables for which we already know the response variable, while the test set is used to validate the accuracy of the model, the generalization set is the one for which the redshift is unknown and the ML algorithm is applied for inferring this information.
First, we use LASSO and select important features based on the data from the training set. Then,
we construct the prediction model using the Superlearner ensemble algorithm which includes the optimized XGBoost, Random Forest, Bayes GLM, and Big LASSO. 
In our case, since the test set has never been used in the training set, then it is called validation data set. 
}

SuperLearner \citep{van2007super} is an algorithm that utilizes 
{k-fold CV to estimate the performance of ML algorithms. It creates an optimal weighted average of the input models, i.e., an ensemble. Namely, the SuperLearner provides coefficients that reflect the relative importance of each learner against the others in the ensemble. Besides this feature, Superlearner can test the predictive power of multiple ML models or the same model, but with different settings. The weights of the algorithms always sum up to 1 and are always equal to or greater than 0. Using these coefficients, we can group the highest weighted algorithms into an ensemble and improve the prediction more than any single algorithm \citep{polley2010super}}. 

We use the functions implemented in the statistical software R, particularly the SuperLearner package.

{In 10fCV the dataset is randomly partitioned into 10 complementary subsets. The SuperLearner is trained on 9 of these subsets and the resulting model is employed to infer the values in the remaining subset, which plays the role of the test set. The process is iterated 10 times, with each subset playing the role of the test set. The SuperLearner parameters are automatically set to optimize the prediction for all test sets (i.e., all data points). Following statistical practice, we repeat this whole procedure 100 times to make the prediction less dependent on the selection of the specific random partition of the dataset. Thus, our predictions result as the average of 100 independent SuperLearner predictions. This allows for stabilization and de-randomization of our results.
Given the paucity of our dataset, this is a crucial step in analyzing the performance of our model.
} 
\section{\textbf{Results}} \label{sec:results}
{Our final training set consists of 657 $\gamma$-ray AGNs with observed redshifts. We separate 73 $\gamma$-ray AGNs as a validation set that is not used for any training (see Fig. \ref{fig:flowchart}). 
}

\begin{figure}[H]
    \centering
    \includegraphics[width=0.49\textwidth]{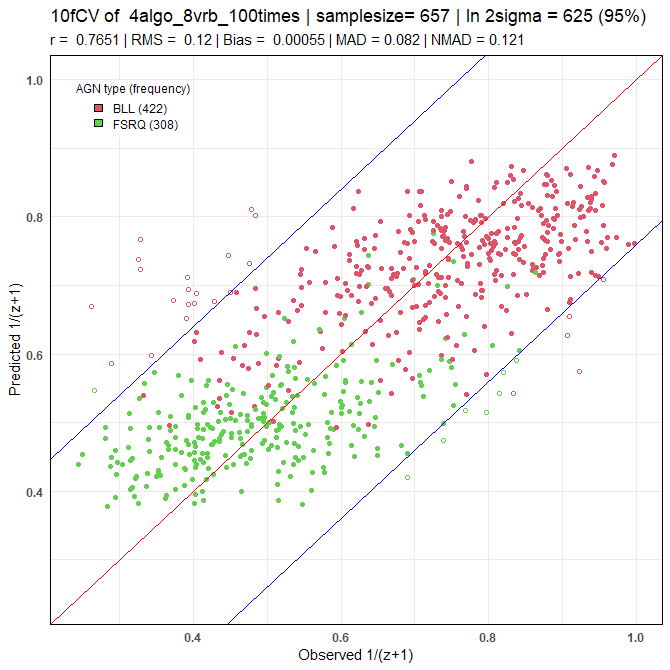}
    \includegraphics[width=0.49\textwidth]{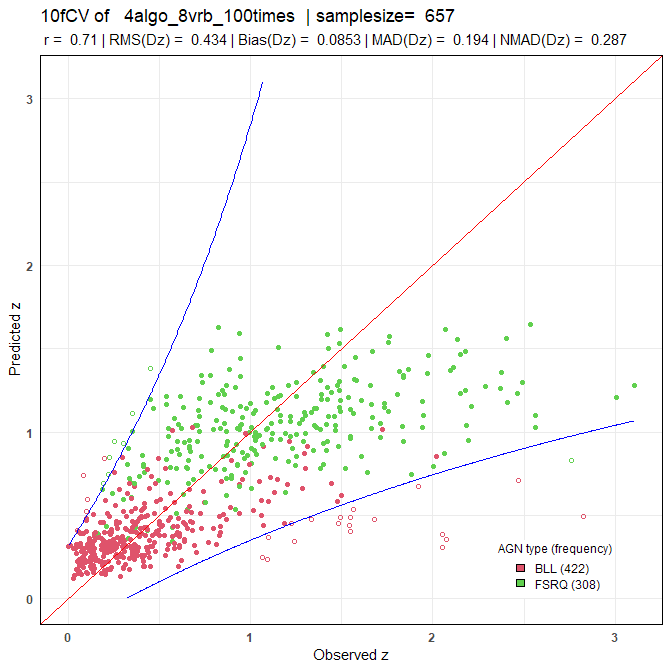}

    \caption{The left panel shows the observed vs. predicted redshift in the $\frac{1}{z+1}$ scale, while the right panel shows the observed vs. predicted redshifts in the linear scale.
    }
    \label{fig:O1}
\end{figure}

\begin{figure}
    \centering

    \includegraphics[width=0.49\textwidth]{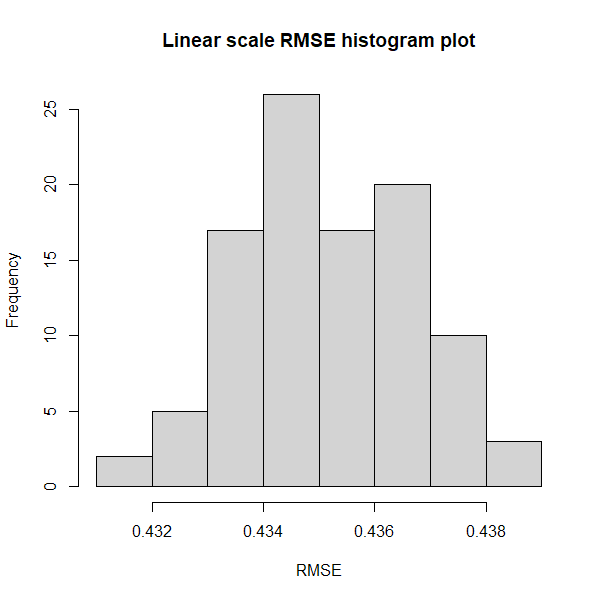}
    \includegraphics[width=0.49\textwidth]{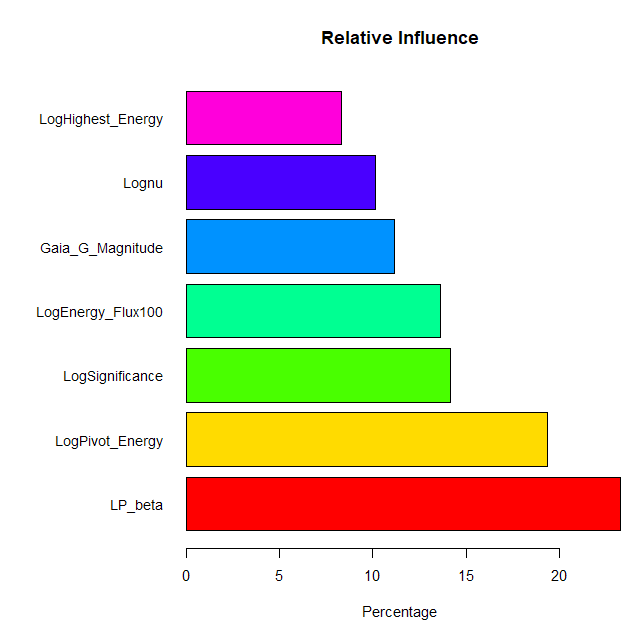}
    \includegraphics[width=0.49\textwidth]{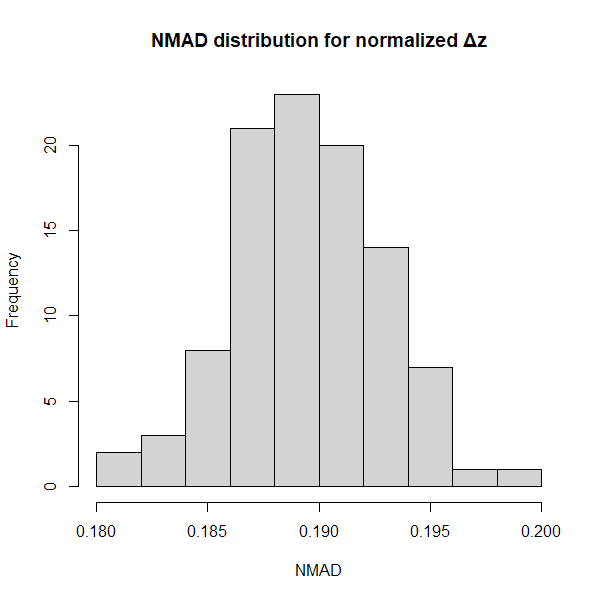}
    \includegraphics[width=0.49\textwidth]{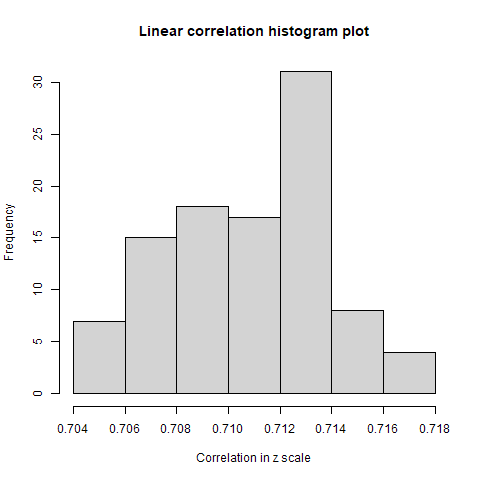}
    
    \caption{In all panels the results are obtained with the one hundred 10fCV. Top left and right panels show the histogram of the RMSE, and the relative influence of our chosen predictors, respectively. Bottom left and right panels show the NMAD distribution, and linear Correlation distribution, respectively.
    }
    \label{fig:histoResults}
\end{figure}

\begin{figure}[H]
    \centering

    \includegraphics[width=0.49\textwidth]{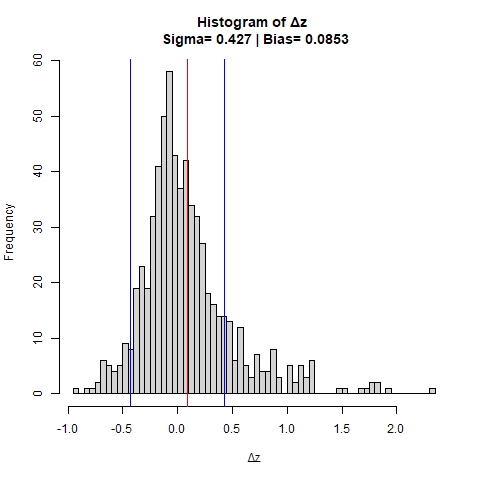}
    \includegraphics[width=0.49\textwidth]{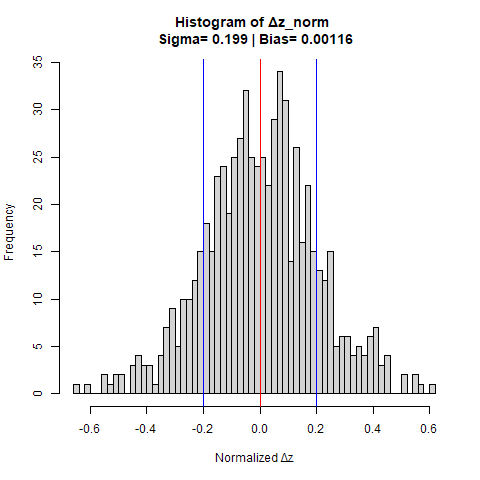}
    \includegraphics[width=0.49\textwidth]{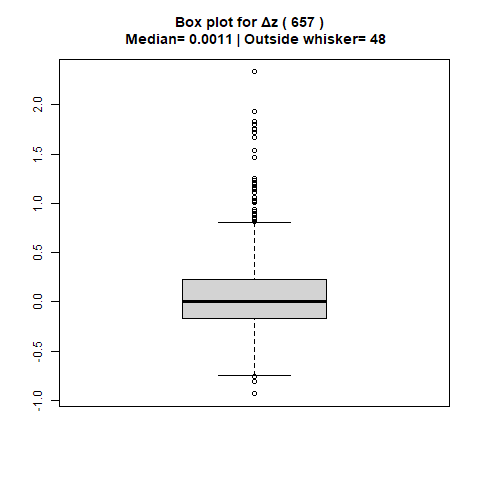}
    \includegraphics[width=0.49\textwidth]{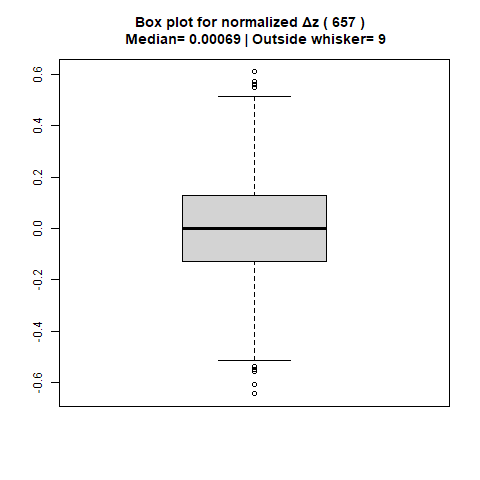}
    
    \caption{The differential distribution of the frequencies $\Delta z$ and $\Delta z_{norm}$ are shown in the left and right panels, respectively. The blue lines indicate the $\sigma$ value and the red line the bias. The bottom plots show the box plot representation of the above frequency histogram, respectively.}
    \label{fig:zspread}
\end{figure}

\begin{figure}[H]
    \centering

  \includegraphics[width=0.49\textwidth]{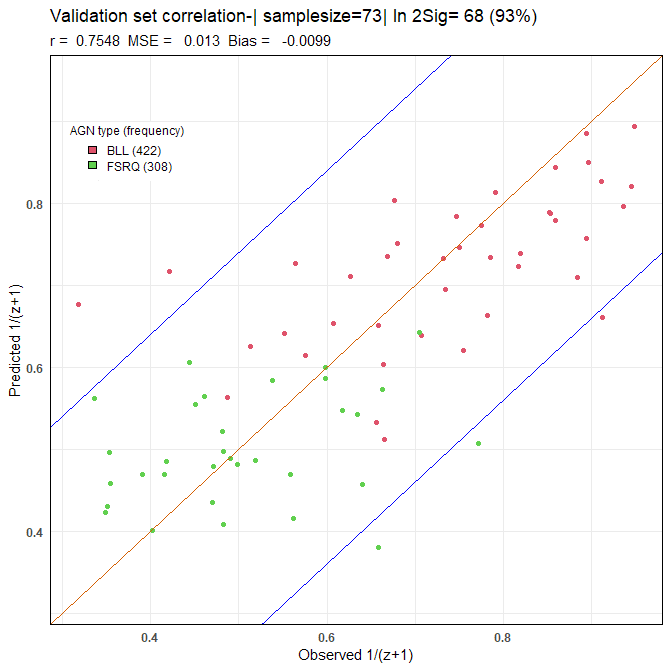}
  \includegraphics[width=0.49\textwidth]{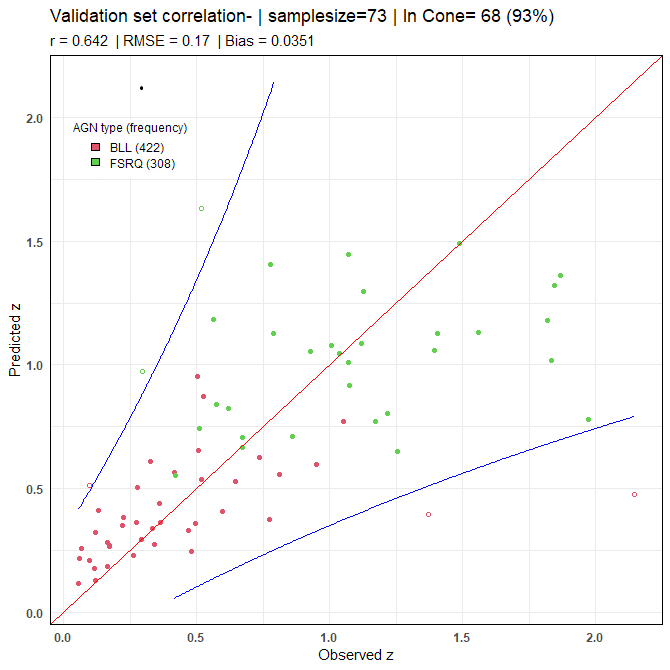}
  
  \caption{{ The Correlation of validation set predicted $\frac{1}{z+1}$ vs. the observed $\frac{1}{z+1}$ (upper left panel) and the predicted z vs. the observed one (upper right panel).}} 
    \label{fig:testset}
\end{figure}

{In Fig. \ref{fig:O1} the top panel shows the correlation plot between the observed and predicted redshift in $\frac{1}{z+1}$ (left panel) and linear scale (right panel). The blue lines indicate the 2$\sigma$ cones for each of the plots where the $\sigma$ is calculated in the $\frac{1}{z+1}$ scale as follows:}
$$ \frac{1}{z_p + 1} = \frac{1}{z_s + 1} \pm 2\sigma, $$
\\
where $z_s$ is the spectroscopic redshift and $z_p$ is the photometric redshift.
{Due to the choice of our scaling, the 2$\sigma$ line is not straight on the linear scale and is shown in the following formula-}
$$
z_p = z_s \left[ \frac{1 \pm 2\sigma(z_p + 1)}{1 \mp 2\sigma} \right] \pm \frac{2\sigma}{1 \mp 2\sigma.} 
$$
{We obtain a Pearson Correlation $r$= 0.71 in the linear scale, with the $\sigma_{NMAD}(\Delta z_{norm})$ = 0.192 and $\sigma_{NMAD}(\Delta z)$ = 0.287. We obtain a low bias for $\Delta z_{norm}$ at $11.6 \times 10^{-4} $ and for $\Delta z$ at $8.5\times10^{-2}$.
We also have a low percentage of catastrophic outliers at 5\% of our total sample. The so-called catastrophic outliers are the outliers in ML nomenclature \citep{jones2020tests}). More specifically, these catastrophic outliers are the $\gamma$-ray AGNs for which $|{\Delta z}| > 2\sigma$, and thus lie outside the cone presented in Fig. \ref{fig:O1}.
In the upper panel of Fig. \ref{fig:histoResults}, we present the distribution of our linear scale RMSE, and the relative influence of the features in our data over the one hundred 10f nested CV runs, in the upper left and right panels, respectively. In the bottom panel of Fig. \ref{fig:histoResults}, the NMAD and the differential distribution of the correlation coefficient are shown in the left and right panels, respectively.
\textcolor{black}{We note here that in our analysis the redshift of $\gamma$-ray AGNs is not just an effect of distance-brightness relation, which is due to selection biases (see \cite{singal2013cosmological}, \cite{singal2012flux}, \cite{singal2014gamma}, \cite{singal2015determination}, \cite{singal2013flat}, as we have discussed in the introduction). 
 Indeed, a very recent study (\cite{qu2019gamma} and \cite{zeng2021cosmological}) has been performed on the 4LAC catalog to evaluate the dependence of the BLLs luminosity on the redshift.
For completeness, we also present the results from a sample that is not used in the CV step at all, alongside with the prediction of the model on an internal test set in Fig. \ref{fig:testset}.}
}
{With this validation set, we have a catastrophic outlier percentage of 7\%, thus comparable with the previous values. 
In the left upper panel of Fig. \ref{fig:zspread}, we show the histogram of $\Delta$z indicating with the red line indicating the bias and with the blue line the $\pm 1$ $\sigma$; 
while in the right upper panel of Fig. \ref{fig:zspread} we present the histogram of $\Delta z_{norm}$ with the red line the normalized bias, and with the blue line the $\pm1 \sigma$ normalized.}

\begin{figure}[H]
    \centering
    \includegraphics[width=0.48\textwidth]{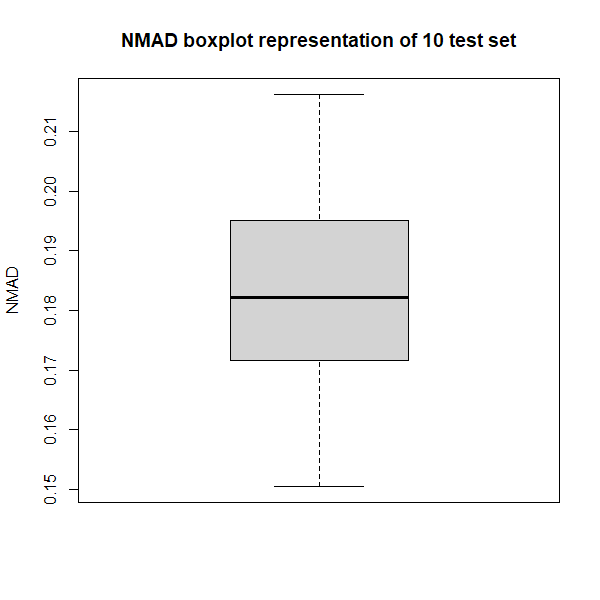}
    \includegraphics[width=0.48\textwidth]{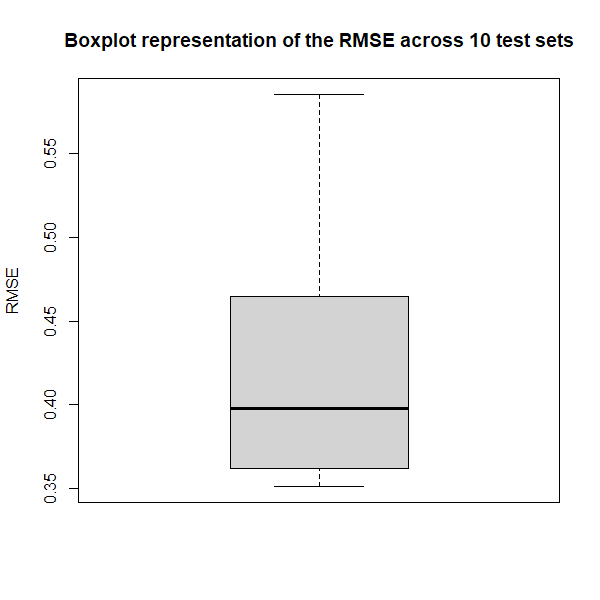}
    \includegraphics[width=0.49\textwidth]{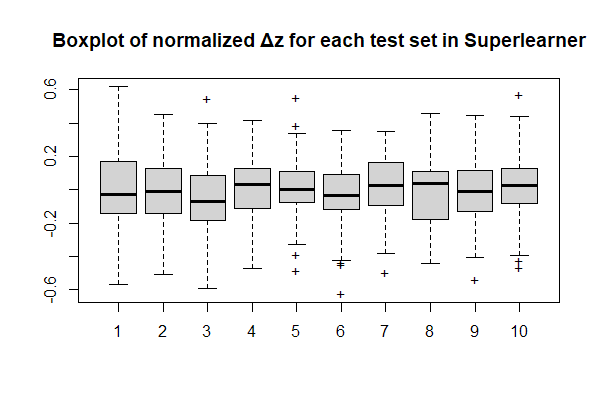}
    \includegraphics[width=0.49\textwidth]{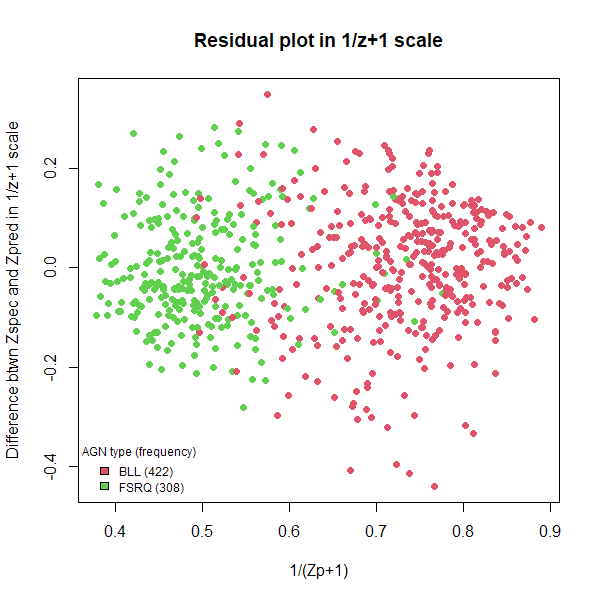}
    
    \caption{The top left and right panels show the boxplot of the NMAD values for the internal Superlearner test set, and the boxplot of the RMSE values for ten internal Superlearner test sets, respectively.
    The bottom left panel shows the $\Delta$z distribution for the ten Superlearner test sets.
    \textcolor{black}{Bottom right panel: The residuals VS the Superlearner predictions for each of the test sets.} }
    \label{fig:residual}
\end{figure}

{We present the residual plot in Fig. \ref{fig:residual} bottom right panel. The lack of any increasing or decreasing trend of the redshift between the residuals and the fitted values is evidence of the goodness of our fit. Furthermore, the $R^2$ value for our result is 0.508, and the (Interquartile Range, IQR ) value for $\Delta z$ = 0.39.}
{Additionally, we compare our results with other works done in the field, such as 
\cite{Richards_2008} (Type-1 broad line quasars from SDSS),
\cite{Laurino_2011} (Optical galaxies and quasars from SDSS ),
\cite{Ball_2008} (Main sample galaxies, luminous red galaxies and quasars from SDSS and GALEX), and 
\cite{brescia2013photometric} (Quasars from SDSS+GALEX+WISE+UKDISS). The comparisons are shown in Table \ref{tab:table1}.}

\begin{table}[]
    \centering
    \begin{tabular}{c|c|c|c}
    \hline
        \textbf{Experiment} & \textbf{Bias} ($\Delta z_{norm}$) & \textbf{Sigma} ($\Delta z_{norm}$) & \textbf{NMAD} ($\Delta z_{norm}$) \\
    \hline    
        Superlearner & 0.001 & 0.19 & 0.19 \\ 
    \hline
        Brescia et al. 2013 (best case) & 0.004 & 0.069 & 0.029 \\
        Laurino et al. & 0.095 & 0.16 & ...\\
        Ball et al. & 0.095 & 0.18 & ...\\
        Richards et al. & 0.115 & 0.28 & ...\\
    \end{tabular}
    \caption{Comparison of our results with those of other ML-based photometric redshift estimation techniques. The empty spaces indicate a lack of available data for those cases.}
    \label{tab:table1}
\end{table}

{We stress that even though our results do not always achieve a more precise prediction than some of the cases shown in Table \ref{tab:table1}, they are still comparable to them, and we need to take into account that our training set is at least twice smaller compared to the sample investigated in the mentioned paper.
Hence, these results highlight that further enlargement and enhancements to the 4LAC dat will produce more precise results in the near future.}


\begin{figure}[H]
    \centering

     \includegraphics[width=0.49\textwidth]{Final Figures/10fCV_plot_4algo_8vrb_100times.png}
    \includegraphics[width=0.49\textwidth]{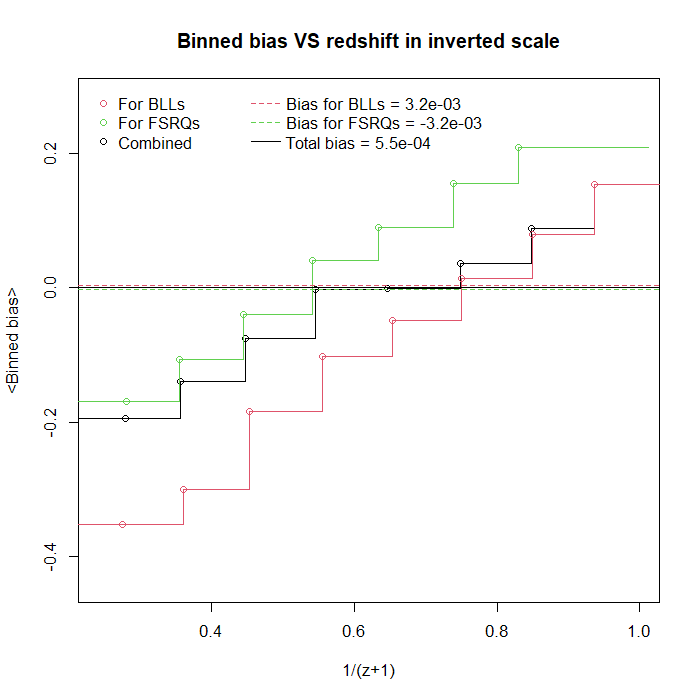}
    
    \caption{Left Panel: Linear regression fitting between the predicted and observed redshifts. The cyan and purple lines show the linear fit for BLL and FSRQs, respectively.
    Right Panel: Plot of binned $\frac{1}{z+1}$ vs mean bias. The average bias for BLLs and FSRQs is $3.2\times10^{-3}$ and $-3.2\times10^{-3}$, respectively.}
    \label{fig:biasfit}
\end{figure}

\textbf{\subsection{Bias correction}}
{As it can be seen from Fig. \ref{fig:O1} left panel, the higher redshift AGNs are being predicted at a lower value. This is a clear signature of our predictions being biased. To correct for this, we fit a linear model between the 
observed and predicted redshifts in the $\frac{1}{z+1}$ scale. We fit linear models for both BLLs and FSRQs separately, which are shown by the cyan and purple dashed lines in Fig. \ref{fig:biasfit}, left panel.
The black dotted line represents the linear fit for both BLLs and FSRQs together. We can see clearly that the fitted lines deviate from the 1:1 line.}

{The bias corrections for BLLs and FSRQs follow this equation:}
\begin{equation}
    U_{prediction} = a * U_{observed} + b, 
\end{equation}

{where $U_{prediction} = \frac{1}{Z_{predictions}+1}$, $U_{observed} = \frac{1}{Z_{observed}+1}$, $a$ and $b$ are the slope and the intercept of linear fit, respectively. We obtain a different value of $a$ and $b$ for BLLs and FSRQs.
These quantify the bias present in our analysis. 
For BLLs : $a$ = 0.29 and $b$ = 0.51. 
For FSRQs : $a$ =   0.29 and $b$ = 0.35.}

\subsection{Prediction on the generalization set}

{Our initial aim, as already indicated in the introduction, is to increase the number of 4LAC $\gamma$-ray AGNs that have estimates of the redshift.
Based on the results shown in the previous section, we have reached so far a trained model which enables predictions for 4LAC $\gamma$-ray AGNs that fall within its trained parameter space. Indeed, for the generalization set, it is of crucial importance to ensure that the generalization set parameter space should overlap with our training set as much as possible.
We start with a great advantage with this data set, since based on the scatter matrix plot in Fig. \ref{fig:Scatter} we can observe that there is a significant overlap in the training (red and green data points for BLLs and FSRQs, respectively) and the generalization set (blue and black points for BLLs and FSRQ, respectively). 
Hence, the trained model has the advantage of extrapolating less when predicting the redshift of the generalization set. For the generalization set, we decide to retain $\gamma$-ray AGNs based on the condition that the values of their predictors should fall within the maximum and minimum values of the corresponding predictor in the training set. This way, we can achieve more reliable redshift predictions with minimal extrapolation.} 
 
{To better evaluate how the generalization set overlaps with the training set, we present a scatter matrix plot in Fig. \ref{fig:overlap}, showing the distribution of the very same seven predictors chosen by the LASSO features in Fig. \ref{fig:Scatter}. The blue points belong to the new trimmed generalization set, and as we can see, all the points fall well within the training set data points, as shown by the red points.}

\begin{figure}[H]
    \centering

    \includegraphics[width=0.84\textwidth]{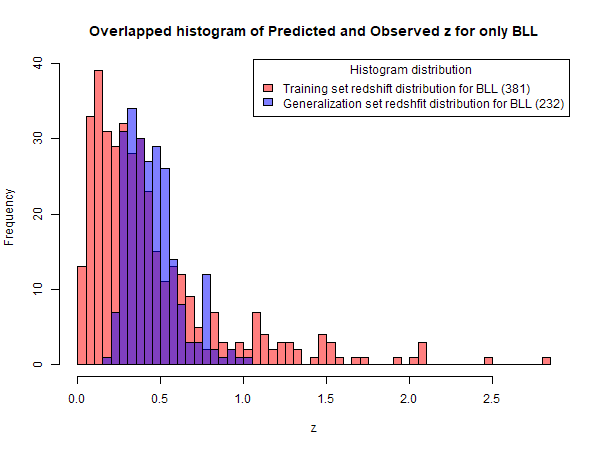}
    \includegraphics[width=0.84\textwidth]{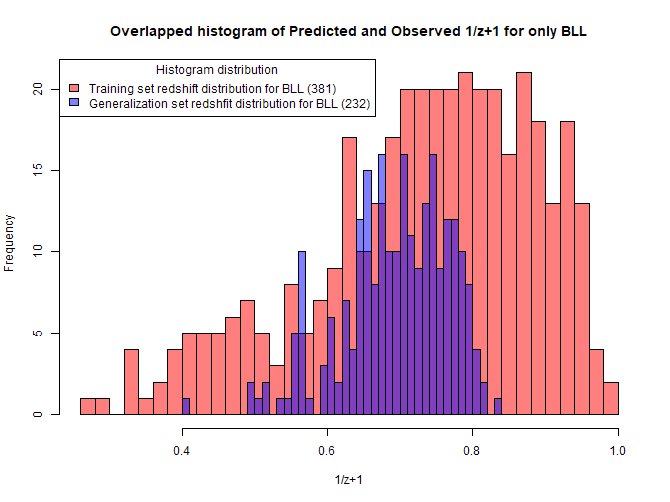}
    \caption{The differential distribution of the predicted redshift of 232 BLLs from the generalization set (blue histogram) vs. training set (orange data points).
    The upper panel shows the distribution in linear scale, while the bottom panels shows the distribution in the $\frac{1}{z+1}$ scale.
    }
   \label{fig:prediction}
\end{figure}

{After we perform these cuts in the parameter space, we are left with 232 $\gamma$-ray AGN which is 97\% of the total number.
These 232 $\gamma$-ray AGNs are all BLLs. 
We would like to clarify here that the objects in the generalization sample that are classified as BCU, or uncategorized, are excluded when we are performing our predictions. We also exclude the single FSRQ that we have in our generalization set, so as to focus solely on BLLs for our predictions.
Thus, the trimming of the variables does not influence the total number of redshifts we predict.
We present the results of our analysis in Fig. \ref{fig:prediction}. As shown in our previous results (see Fig. \ref{fig:O1}), 95\% of our predictions fall within the 2$\sigma$ error bars. We expect a similar scenario for the predictions on the generalization set. Here, the blue histogram bars represent the median of the predictions on the generalization set, not taking into account the 2$\sigma$ errors.
We performed the Kolmogorov Smirnov Test (KS) test to evaluate if the extracted redshift distribution comes from the observed redshift distribution in the training set. As a result, we obtained that the null hypothesis that the two distributions come from the same parent population is rejected at the level of less than $10^{-16}$\%.  Since we are not taking into account the error bars, hence the KS test gives us that the two distributions are different. 
Thus, we decided to investigate this issue by performing the KS test again on the singular distribution of the variables and we confirm also that the null hypothesis of similarity is rejected. 
Thus, it is not surprising that the two redshift distributions are not similar. Nevertheless, we do not necessarily expect the distributions of the redshift to be similar from a statistical point of view, since selection biases are at play and it is possible, as mentioned earlier, that we observe the faintest $\gamma$-ray AGNs at low redshift and the brightest $\gamma$-ray AGNs at higher redshift.
}

{
Our model without accounting for the bias correction predicts the redshift for BLLs between 0.5 and 1. With the application of the bias correction, the predicted redshifts are extended to cover the whole interval between 0 and 3, which better resembles the distribution of true redshifts.
When the originally predicted redshift (Superlearner prediction) is close to 0.5, then we are at the borders of the generalization limits, namely close to the intercept values $b$ and can not predict the true redshift well.
}

{
{To be more specific, our sample contains FSRQs and BL Lacs in similar numbers (655 FSRQs and 686 BL Lacs). However, it is easier to measure redshift in FSRQs given their prominent broad emission lines. Given the observational difficulties in measuring redshifts for BL Lacs, the sources in our study might not be a representative sample of the BL Lac population. There is a non-zero probability chance for sources to be mis-classified, or even the $\gamma$-ray source to be mis-associated with a counterpart. Moreover, our sample contains only 60 non-blazar $\gamma$-ray AGN whose $\gamma$-ray properties potentially evolve differently with redshift.  All of the above may hamper the accuracy of the ML models. However, given the improvement in localization accuracy, the number of sources, and the number of non-blazar $\gamma$-ray AGN (a factor of two improvement) between the 3LAC and 4LAC (as well as earlier catalogs), future {\it Fermi} catalogs will allow us to address further the shortcomings of our current sample.}

\section{\textbf{Conclusion}} \label{sec:conclusion}
In this work, we have crafted a methodology to predict the redshift of $\gamma$-ray loud AGN from the 4LAC catalog, using their observed $\gamma$-ray properties.
{We used categorical variables to distinguish among $\gamma$-ray AGN types and the LASSO algorithm to select the most predictive variables. We select the ML models based on the coefficient of the predictive power obtained with Superlearner after we have performed the optimization of the models.}
We trained several ML algorithms on these properties by using Superlearner and used the trained models to predict the $\gamma$-ray AGN redshifts. 
By computing the relative influence of these observed properties, we also determine which of them are the best predictors. 
The application of these methods to the 4LAC $\gamma$-ray AGN catalog for the BLLs sources for which the redshift is unknown increases 61\% the size of the data set of $\gamma$-ray AGNs with known redshift, thus allowing to reach a larger sample. This new data set will have the great advantage to be complete for a given flux limit with a higher percentage.
This enlarged sample of $\gamma$-ray AGNs, in turn, will allow us to determine the luminosity function, its evolution, and the density evolution of $\gamma$-ray AGNs with improved accuracy. 
With a sample of 657 $\gamma$-ray AGNs with measured redshifts, we have shown that using the Superlearner method can provide predicted redshifts that correlate with the observed redshift to a high degree of accuracy. 
{We obtain, after performing one hundred 10f nested CV, an average Pearson Correlation coefficient, $r=0.77$ in the $\frac{1}{z+1}$ scale and RMSE$ = 0.12$ and a bias of $5.4 \times 10^{-4}$; if we consider the results instead in the z scale $r=0.71$, the RMSE$(\Delta z_{norm})=0.43$, the bias$(\Delta z_{norm})$ $1.2 \times 10^{-3}$ and $\sigma_{NMAD}= 0.192$.
 }

We then predict the redshift of 232 BLLs that do not have the observed redshift and plot them against the observed redshift. Most $\gamma$-ray AGNs without the estimation of redshift lie between $0.18 \leq z \leq 1.02$.

Previous work utilizing ML algorithms focused primarily on the classification of $\gamma$-ray AGNs. 
Currently, to the best of our knowledge, no work in the {\bf blazar} literature attempts to estimate the redshift using their observed $\gamma$-ray characteristics. This is a pioneering work in $\gamma$-ray AGN redshift estimation and will hopefully usher in follow-up studies that can improve our predictive capabilities even further. 


\section{Appendix}
{In this Appendix, we discuss how it is crucially important to show how the models used together with an ensemble performs better than the singular methods.}
{In Table. \ref{tab:table2} we show the RMSE, linear correlation, Bias, and NMAD scores of the individual algorithms used in the ensemble and the final Superlearner ensemble score. 
Based on the RMSE and the linear correlation values, we can clearly see that the Superlearner ensemble performs better.} 
{The singular model scores presented here are 10fCV and we ran them with the same optimization parameters shown in Sec. \ref{sec:optimizing}.}

}

\begin{table}[H]
    \centering
    \begin{tabular}{c|c|c|c|c}
    \hline
    \textbf{Algorithm} & \textbf{Root Mean Square error} & \textbf{Linear correlation} & \textbf{Bias ($\Delta z_{norm}$) ($\times10^{-4}$)} & \textbf{NMAD $\Delta z_{norm}$} \\
    \hline
        SuperLearner & 0.014 & 0.71 & 11.6 & 0.19 \\
    \hline
        XGB & 0.015 & 0.70 & 22.6 & 0.19 \\
        RF & 0.015 & 0.70 & 15 & 0.20 \\
        BigLasso & 0.02 & 0.69 & 2.2 & 0.19 \\
        BayesGLM & 0.02 & 0.69 & 8.6 & 0.19 \\
    \hline
    \end{tabular}
    \caption{The 10fCV risk estimates of individual algorithms and the Superlearner ensemble. }
    \label{tab:table2}
\end{table}


{Our choice of using $\frac{1}{z+1}$ scaling for the redshift instead of log(z+1) is based on the result presented in Table. \ref{tab:table3}. These results are obtained after performing a 10fCV using the two different scalings.}

\begin{table}[H]
    \centering
    \begin{tabular}{c|c|c|c|c}
    \hline
        \textbf{Scaling} & \textbf{Mean square error} & \textbf{Linear Correlation} & \textbf{Bias ($\Delta z_{norm}$) ($\times10^{-4}$)} & \textbf{NMAD $\Delta z_{norm}$} \\
        \hline
        log(z+1) & 0.427 & 0.70 & 223 & 0.2 \\
        $\frac{1}{z+1}$ & 0.435 & 0.71 & 11.6 & 0.19 \\
    \hline
    \end{tabular}
    \caption{The MSE, Correlation, bias, and NMAD of two different redshift scaling. }
    \label{tab:table3}
\end{table}

\begin{figure}[H]
    \centering

    \includegraphics[width=0.75\textwidth]{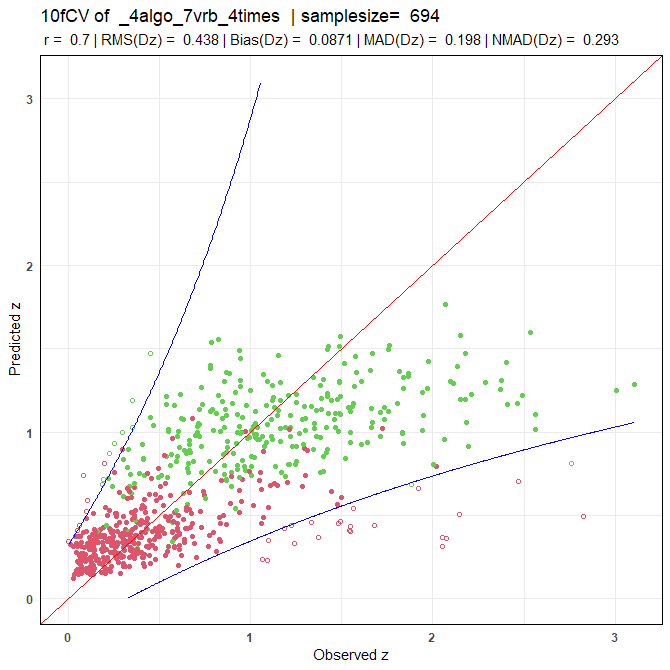}
    
    \caption{The linear scale correlation plot, when LogSignificance is not included.}
    \label{fig:woLS}
\end{figure}

We show the one hundred 10fCV results related to the RMSE, the NMAD distribution for the normalized $\Delta z$, and the linear correlation. 
For completeness of the discussion, we show the results when we exclude LogSignificance from our analysis, see Fig. \ref{fig:woLS} .

\begin{figure}[H]
    \centering

    \includegraphics[width=0.75\textwidth]{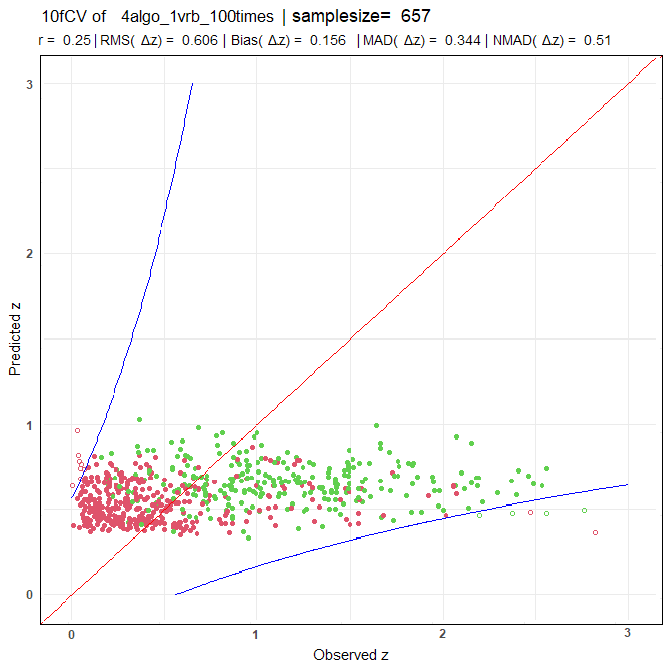}
    
    \caption{Correlation plot in linear scale. The values for statistical parameters are shown on the plots themselves.}
    \label{fig:logef}
\end{figure}

Next, we present the results when we use only a single variable, LogEnergyFlux, for the prediction using our ensemble, in Fig.\ref{fig:logef}.

{It is clear that when we use only one predictor even though it has a high relative influence (the flux), the prediction we achieve for the redshift is poor compared to the prediction we obtain with the full set of LASSO selected predictors.}

\begin{figure}[H]
    \centering

    \includegraphics[width=0.75\textwidth]{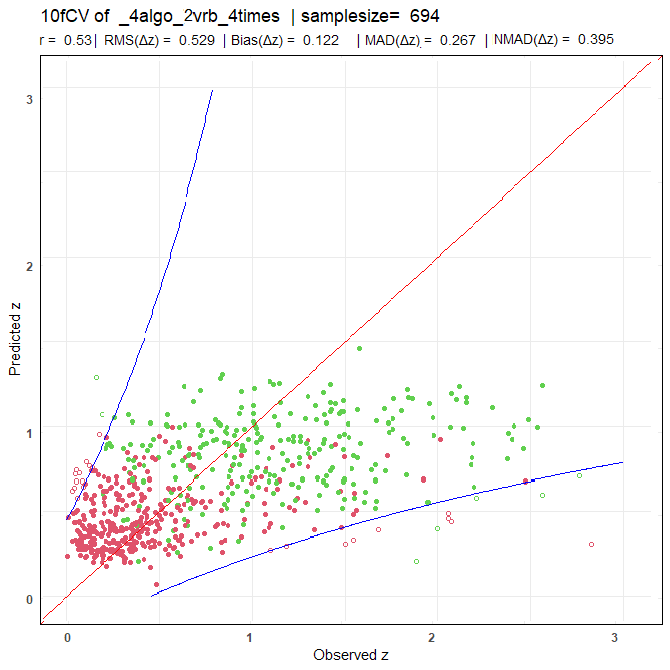}
    
    \caption{The linear correlation plot when using LP\_beta and LogPivotEnergy in our ensemble. }
    \label{fig:high2corr}
\end{figure}

{Additionally, we show the results obtained when using our two most predictive features, i.e LP\_beta and LogPivotEnergy in Fig. \ref{fig:high2corr}.}

{These two have the highest relative influence in our feature set, but, using them independently does not lead to accurate results as the entire feature set does.}

\acknowledgments
 This work presents results from the European Space Agency (ESA) space mission, Gaia. Gaia data are being processed by the Gaia Data Processing and Analysis Consortium (DPAC). Funding for the DPAC is provided by national institutions, in particular the institutions participating in the Gaia MultiLateral Agreement (MLA). The Gaia mission website is https://www.cosmos.esa.int/gaia. The Gaia archive website is https://archives.esac.esa.int/gaia.
M.G.D. thanks Trevor Hastie for the interesting discussion on overfitting problems.
We also thank Raymond Wayne for the initial computation and discussions about balanced sampling techniques which will be implemented in subsequent papers. 
This research was supported by the Polish National Science Centre
grant UMO-2018/30/M/ST9/00757 and by Polish Ministry of Science and
Higher Education grant DIR/WK/2018/12.
Finally, we would like to thank Sarthak Das and Subham Kedia for their help in performing the 10-fold CV analysis of the Superlearner algorithm.

\bibliography{refs}

\end{document}